\documentclass[12pt]{article}
\usepackage{amssymb,bbm}
\usepackage{amsfonts}
\usepackage{latexsym}
\usepackage{amsmath}
\usepackage{amsthm}
\usepackage{geometry} 

\def\lddots{\mathinner{\mkern1mu\raise1pt\hbox{.}\mkern2mu  
\raise4pt\hbox{.}\mkern2mu\raise7pt\vbox{\kern7pt\hbox{.}}\mkern1mu}}
\makeatletter
\def\numberbysection{\@addtoreset{equation}{section}
 \def\theequation{\thesection.\arabic{equation}}}
 
\makeatother  
  
\numberbysection

\newcommand{\be}{\begin{eqnarray}}  
\newcommand{\ee}{\end{eqnarray}} 
\newcommand{\nn}{\nonumber}

\def\ds{\displaystyle}

\def\mf{\mathfrak}
\def\bb{\mathbbm}
\def\C{\bb C}

\def\1{\bb{1}}

\def\pd{\partial}
\def\a{\alpha}

\def\d{\delta}

\def\e{\epsilon}

\def\l{\lambda}

\def\r{\rho}
\def\s{\sigma}

\def\g{\gamma}
\def\G{\Gamma}

\def\bp{\begin{proof}[{\bf Proof}]}

\newcounter{prop}
\newcommand\Prop{{ \vskip .25cm}{\bf Property \theprop}{\addtocounter{prop}{1}}\\}

\begin{document}  
 \begin{center}  
  \LARGE Classification of the solutions of constant rational semi-dynamical reflection equations \\[0.8in]  
\Large  {\it Dedicated to Daniel Arnaudon}\\[0.8in] 
\large  { Jean Avan} \footnote{e-mail: avan@u-cergy.fr}
and
\large  { Genevi\`eve Rollet} \footnote{e-mail: rollet@u-cergy.fr}\\  
\normalsize{{Laboratoire de Physique Th\'eorique et Mod\'elisation\\ Universit\'e de  Cergy-Pontoise,  
5 mail Gay--Lussac, Neuville--sur--Oise,\\F-95031 Cergy--Pontoise Cedex}}\\  
\end{center}

\begin{abstract}  
We propose a classification of the solutions $K$ to the semi-dynamical reflection equation with constant rational structure matrices associated to rational scalar Ruijsenaars-Schneider model.
Four sets of solutions are identified and simple analytic transformations generate all solutions from these sets.
\end{abstract}  

\section{Introduction}  
Reflection equations appeared in the factorized scattering on a half-line~\cite{cherednik, sklyanin}; they describe consistency conditions between reflection matrices $K$ and bulk scattering matrices $R$ guaranteeing  integrability of quantum systems with non-periodic boundary conditions. 
These equations take the form:
\be R_{12}\ K_{1}\ \tilde{R}_{21}\ K_{2} =  
K_{2}\ \tilde{R}_{12}\ K_{1}\ R_{21}. \label{refl} \ee 
As usual, the auxiliary spaces indexed by $(1,2)$ may be loop spaces $V\otimes \C(\l)$ and the spectral parameter dependence of $R$ and $K$ is then implicit.

General quadratic exchange algebras were soon considered~\cite{KuSa,FrMa} stemming in particular from the study of quantum integrable non-ultra-local field theories~\cite{MAI2}.
Their general form reads:
\be A_{12}\ K_{1}\ B_{12}\ K_{2} =  
K_{2}\ C_{12}\ K_{1}\ D_{12}, \label{quad} \ee 
with consistency conditions: unitarity of $A$ and $D$, 
$C_{12}=B_{21}$, Yang-Baxter type cubic equations for $A, B, C, D$.

They have been the object of many studies, running from their covariance properties~\cite{KuSa}, their interpretation as twists of quantum groups (non-affine)~\cite{KDM} to the classification of solutions $K$ for a variety of $A, B, C, D$ matrices (see~\cite{KuSa, LiSan, Ku, AACDFR,MENE}) to the use of explicit $K$-solutions to build integrable spin chains with integrable boundary (see~\cite{sklyanin}) and the study of their symmetry properties~\cite{DM}.

Similarly to the Yang-Baxter equation, the generalized reflection equations admit ``dynamical'' extensions.
Precisely, the structure matrices $A, B, C, D$ and the reflection matrix $K$ then depend on parameters $\{\l_i\}$ interpreted as coordinates on the dual of a Lie Algebra $\mf{h}$ (often abelian, although the notion of non-abelian quantum dynamical group exists, see e.g.~\cite{Ping,DMu, Et}).
The consistency equations for $A, B, C, D$ and the quadratic equation for $K$ now occur with consistent shifts on these parameters, and zero-weight compatibility conditions under adjoint action of $\mf{h}$ on $A, B, C, D$.
Contrary to the quantum-group  case, where only one extension is identified (Gervais-Neveu-Felder equation, see~\cite{GFN,ABB}), at this time two unequivalent dynamical extensions of~(\ref{quad}) have been found.

The first one, known as either ``boundary dynamical quantum group''~\cite{FHS} or ``fully dynamical'', can be deduced from~(\ref{quad}) by the famous vertex-IRF transformation~\cite{Ba}.
It has been studied quite extensively, with construction of solutions~\cite{FHS}, interpretation as twist~\cite{KuMu}, construction of quantum integrable systems \`a la Gaudin~\cite{YSa}; recently the general form of quantum traces, and local spin chains associated to it, was described~\cite{NA3}.
The equation reads:
\be A_{12}\ K_{1}(\l+h_2)\ B_{12}\ K_{2}(\l+h_1) =  
K_{2}(\l+h_1)\ C_{12}\ K_{1}(\l+h_2)\ D_{12}, \label{fulldyn} \ee 
with zero weight conditions for the four structure matrices ($A_{12}, B_{12}, C_{12}$ and $D_{12}$ commute with $h_1+h_2$).

The second dynamical extension of~(\ref{quad}), which we will consider here, has been characterized only recently~\cite{NA1}.
Originally found in the quantization of the scalar Ruijsenaars-Schneider $r$-matrix structure~\cite{ACF}, it received its general formulation and analysis of comodule and 'coproduct-like' properties in~\cite{NA1}; quantum traces and local spin-chain Hamiltonians were discussed in~\cite{NA3}.
The quadratic exchange relation has the form:
\be A_{12}\ K_{1}\ B_{12}\ K_{2}(\l+h_1) =  
K_{2}\ C_{12}\ K_{1}(\l+h_2)\ D_{12}, \label{semidyn} \ee 
with $C_{12}=B_{21}$ and zero weight conditions for only two structure matrices (precisely $B_{12}$ commutes with $h_1$ and $D_{12}$  with $h_1+h_2$).

Very few examples are known, in fact only the quantization of scalar Ruijsenaars-Schneider model provides at this time explicit forms of $A, B, C, D$ and $K$ matrices.
It is however clear that in order to both  understand the algebraic meaning of that structure, and to use it in explicit constructions of integrable Hamiltonians~\cite{NA3} we must find a general classification of solutions.

More precisely, the problem is twofold:
find $A, B, C, D$ constrained by a set of Y.B.-like equations compatible with the associativity of the algebra;
find $K$ for such $A, B, C, D$'s.
Contrarily to the Y.B. case, the two questions are decoupled.
Note that several distinct forms of YB-type equations may arise as sufficient conditions, noticeably when a spectral parameter dependance occurs~\cite{ACF}.
This will be commented upon in the conclusion.

We shall here restrict ourselves to the ``simplest'' second problem:
for a given set of rational, non-spectral parameter dependent 
(called here "constant" in the usual $R$-matrix terminology)
$A, B, C, D$, find all possible reflection matrices $K$.
The set of $A, B, C, D$ we will start with is the one obtained in~\cite{ACF} describing the Lax matrix structure of quantum rational Ruijsenaars-Schneider model~\cite{RS}.

First of all we shall describe two general properties of equation (\ref{semidyn}), including two possible ways of constructing extra solutions from a given one: a gauge transformation property and a decoupling ``column degeneracy'' property (actually a limit of the first one). 

We shall then identify four sets of meromorphic solutions $K$ exhausting the set of solutions. 
The solution in ~\cite{ACF} is of course recovered, but it is understood in a more general formulation, and in addition three other types of solutions are exhibited, one invertible and two rank one projectors.

We shall finally discuss the interpretation of our new solutions, possible extensions to trigonometric, elliptic cases, constant and non-constant, and come back to the question of finding new  $A, B, C, D$'s.

\section{General properties of constant semi-dynamical reflection equations}  

Let us recall that the  semidynamical quadratic exchange relation without spectral parameter reads:
\be A_{12}\ K_{1}\ B_{12}\ K_{2}(\l+h_1) =  
K_{2}\ B_{21}\ K_{1}(\l+h_2)\ D_{12}, \nonumber\\
\mbox{with} 
\quad [B_{12}, h_1]=0
\quad\mbox{and} 
\quad [D_{12}, h_1+h_2]=0.\qquad\label{resemidyn}
\ee
We shall assume that the representation space $V$ of the matrices $K$ (considering a representation space $V \otimes V$ for  $A, B, C, D$ matrices) is a diagonalizable irreducible module of $\mf{h}$.
As a consequence we shall identify $\mf{h}$ with the Cartan algebra of $gl(n)$ and $V$ with $\C^n$.
The set of equations (\ref{resemidyn}), projected on generators $e_{ij}\otimes e_{kl}$ of $gl(n)$, then takes the form:

\be
\sum_{x y} M^{ijkl}_{xy} K_{xj}\,K_{yl}(\l_j+\g)=
K_{kl}\,\sum_x  N^{ijl}_x K_{xj}(\l_l+\g)\label{csq0w}
\ee
with $M$ and $N$ quadratically depending on $A, B, C, D$ matrix elements.
The notation $f(\l_j+\g)$ will denote the adjoint action of the shift operator $\ds e^{\g\,\pd_j}$~\footnote{$\ds \pd_j={{\pd}\over{\pd \l_j}}$} on any function $f$ of the variables $\l_1,\ldots,\l_n$.

Two general properties of (\ref{resemidyn}) now follow:

\Prop
The set  of equations (\ref{resemidyn}) subdivides into subsets coupling only two fixed columns of $K$ (namely those with indices $j$ and $l$ in (\ref{csq0w})).

\Prop
For a given set of  $A, B, C, D$ matrices, if $K$ is a solution of (\ref{resemidyn})  then for any function $f$ of the variables $\l_1,\ldots,\l_n$, another solution is given by $\tilde{K}$, with 
\be \tilde{K}_{ij}=K_{ij}{{f(\l_j+\g)}\over{f(\l)}}.\label{gauge}\ee
In other words, multiplying a solution $K$ on the right by a diagonal matrix, $\ds \left (f_j(\l)\right )$, satisfying the flatness condition 
\be
f_j(\l)\,f_i(\l_j+\g)=f_i(\l)\,f_j(\l_i+\g),\label{flatcond}
\ee
is a gauge transformation;
the solution of the flatness equations (\ref{flatcond}) being precisely given by a gradient:
\[f_j(\l)={{
f(\l_j+\g)}\over{f(\l)}}.\]
Indeed, the cases where some $f_j$ are equal to zero -perfectly admissible solutions of (\ref{flatcond})- can actually be included in this gradient formulation: $\ds f=\prod_j e^{c_j\,\l_j/\g}$ yields $\ds f_j(\l)={{
f(\l_j+\g)}\over{f(\l)}}= e^{c_j}$ and finally $\ds \lim_{c_j \to-\infty} f_j(\l)=0$.
As a consequence, one can multiply any solution $K$ on the right by a constant diagonal matrix, allowing for instance to set to zero any number of chosen columns.

In the case when the matrix $D$ is not only zero-weight but also ``dynamical zero-weight'' ($[h_1\,\pd_1+h_2\,\pd_2,D_{12}]=0$), equation (\ref{semidyn}) can be rewritten without shifts, albeit with non-abelian $K$-matrices: $\ds \tilde{K}_1=K_1\,e^{\g\,\pd_1}$.
Gauge transformation (\ref{gauge}) can then be interpreted as a canonical transformation redefining $e^{\g\,\pd_j}$ as $\ds f^{-1}(\l)\,e^{\g\,\pd_j}f(\l)$, that is,  $\ds{{f(\l_j+\g)}\over{f(\l)}}\,e^{\g\,\pd_j}$.

\section{Solving the rational case}

In the rational constant case derived in~\cite{ACF} the $A, B, C, D$ matrices read:
\be
A=\1+\sum_{i\ne j}{{\g}\over{\l_{ij}}} \,(e_{ii}-e_{ij})\otimes (e_{jj}-e_{ji})
\nn\\
B=\1+\sum_{i\ne j}{{\g}\over{\l_{ij}-\g}} \,e_{jj}\otimes (e_{ii}-e_{ij})
=C^{\pi}\nn\\
D=\1+\sum_{i\ne j}{{\g}\over{\l_{ij}}} \,(e_{ij}\otimes e_{ji}-e_{ii}\otimes e_{jj})
\label{abcd}
\ee
Equations (\ref{csq0w}) take  different forms, depending on the specific choice of generator $e_{ij}\otimes e_{kl}$. A priori $15$ different cases would have to be considered ($i=j=k=l, i=j=k\ne l$,\ldots, $(i,j,k,l)$ all distinct); in fact, some equations directly reduce to $0=0$,  some equations actually coincide and some are consequences of a set of others. The case $i=j=k=l$ collapses to $0=0$; the set of $7$ cases involving exactly two distinct indices reduces to the following three functional equations:
\be
{{\g}\over{\l_{ij}}} K_{ij}K_{ij}(\l_j+\g)=
K_{ij}K_{jj}(\l_j+\g)
-(1-{{\g}\over{\l_{ij}}}) K_{jj}K_{ij}(\l_j+\g)
\label{2sc}\\
{{\g}\over{\l_{ij}}} K_{ij}K_{ji}(\l_j+\g)=
K_{ij}K_{ii}(\l_j+\g)
-(1+{{\g}\over{\l_{ij}}}) K_{ii}K_{ij}(\l_i+\g)
\label{2sl1}\\
(2-{{\g}\over{\l_{ij}}}) K_{ij}K_{ji}(\l_j+\g)
=K_{ij}K_{ii}(\l_j+\g)
-(1-{{\g}\over{\l_{ij}}}) K_{ii}K_{ij}(\l_i+\g)+\nn\\
(1-{{\l_{ij}}\over{\g}}) (K_{ii}K_{jj}(\l_i+\g)-K_{jj}K_{ii}(\l_j+\g))
\label{2sl2}
\ee
Actually the set of equations (\ref{2sc}) decouples since it's the only one among  the three sets of equations involving a shift on the variable labeled by the column index. 
It describes self-coupling of a single column.
Let us consider this set of equations first.
One immediatly notices that, if the diagonal term $K_{j_0j_0}$ vanishes (as a function), then any other term of the  same column also vanishes\footnote{We reduce our analysis to meromorphic functions.} ($\forall i, K_{ij_0}=0$).
Inserting this into (\ref{csq0w}) yields two types of equations:

-either one with $j$ or $l$ equal to $j_0$, that reduces to $0=0$.

-or one with none of the column indices equal to $j_0$, in this case no term $K_{ij_0}$ is involved in the equation.

This leads us to introduce the set $J_0$ of all indices $j_0$ such that $K_{j_0j_0}=0$. Therefore  the set of equations (\ref{csq0w}) reduces to the subset with no zero-column. We shall now solve this subset of equations, and from now on it will be understood that neither $j$ nor $l$ belong to $J_0$.
Introducing the following natural gauge-invariants:
\be
X_{ij}={{K_{ij}}\over{K_{jj}}};\quad 
R_{jl}={{K_{jj}(\l_l+\g)}\over{K_{jj}}}\,{{K_{ll}}\over{K_{ll}(\l_j+\g)}},
\label{XR}
\ee 
we get:
\be
X_{ij}\, \{
\a_{kj}\,\a_{ik}\,X_{kl}(j)
+(\a_{ik}\,(\a_{ij}-\a_{kj})-\a_{ij}+1)\,X_{jl}(j)
-\a_{ij}\,(\a_{ik}-1)\,X_{il}(j)
 \}\hspace {2cm}
\nn \\
-X_{kj}\,\{
\a_{kj}\,(\a_{ik}-1)\,X_{kl}(j)
+(\a_{ik}\,(\a_{ij}-\a_{kj})+\a_{kj}\hskip 6cm \nn \\
+\a_{lj}\,(1-\a_{ij})-1)\,X_{jl}(j)
+\a_{ij}\,(\a_{lj}-\a_{ik})\,X_{il}(j)
\}
\nn \\=
\a_{lj}\,R_{jl}\,X_{kl}\,\{
\a_{il}\,X_{ij}(l)-(\a_{il}-1)\,X_{lj}(l)
\}\hskip 5cm \label{ijkl}
\ee
introducing here a short-hand notation $(j) \;\mbox{for} \;(\l_j+\g)$
and:
\[\a_{ab}=1+(1-\d_{ab}){{\g}\over{\l_{ab}(\l_j+\g,\l_l+\g})}.
\]
Before actually solving equations (\ref{ijkl}), let us first describe
the reconstruction procedure of $K_{ij}$ from $R_{jl}$ and $X_{ij}$.
It is done in two steps:

\noindent -1- Exhibiting some diagonal part $K_{jj}$ leading to the  found $R_{jl}$.
This requires that $R_{jl}$ obey a zero-curvature condition:
\[
{{R_{ij}(\l_k+\g)}\over{R_{ij}}}\,
{{R_{jk}(\l_i+\g)}\over{R_{jk}}}\,
{{R_{ki}(\l_j+\g)}\over{R_{ki}}}=1
\]
which may \underline{not} be satisfied for some particular solution $R_{jl}$ of (\ref{ijkl}).
In such a case, no associated $K$ exists.
If, on the other hand, some $K_{jj}(\l)$ yield $R_{jl}$, all other solutions for this $R_{jl}$  are given by $\tilde{K}_{jj}(\l)=f_j(\l)\,K_{jj}(\l)$, $\ds \left (f_j(\l)\right )$ being a flat (i.e. satisfying the flatness equations (\ref{flatcond}), hence $\ds f_j(\l)=f(\l_j+\g)/f(\l)$) diagonal matrix,

\noindent -2- Finally $K_{ij}(\l)=X_{ij}(\l)\,K_{jj}(\l)$.

Let us now come back to the resolution of equations (\ref{ijkl}).
We will once again first concentrate on equation (\ref{2sc}) which involves only $X$:
\be
{{\g}\over{\l_{ij}}} X_{ij}X_{ij}(\l_j+\g)=
X_{ij}-(1-{{\g}\over{\l_{ij}}}) X_{ij}(\l_j+\g).
\nn
\ee
It takes the simple form $G_{ij}(\l_j+\g)=G_{ij}$ introducing the invertible parametrization 
$\ds X_{ij}=1+{{\l_{ij}}\over{G_{ij}-\l_{ij}}}$.

The next equation we will take into account is equation (\ref{ijkl}) with $j=l$~\footnote{This requires that $n\ge 3$, we will come back to $n=2$ afterward.}. 
It involves only entries of $G$ in a single column $j$ and with shifts only on $\l_j$, precisely the variable for which $G$ is $\g$-periodic. This equation with three distinct indices actually factorizes in the remarkably nice following way:
\be
(G_{ij}-G_{kj})\,(G_{ij}-G_{kj}-\l_{ik})=0.\label{ijkj}
\ee
The solutions of this last equation read:
\be\ds G_{ij}=G_j+{{1}\over{2}}(1+\e_j)\,\l_i \label{GijGj}\ee where $\e_j$ is a sign and $G_j$ an arbitrary function.
The periodicity condition that $G_{ij}$ has to satisfy is transmitted to $G_j$:
$G_j(\l_j+\g)=G_j$.

To get suitable consistency relations between these $G_j$'s, we now have to consider equations coupling two different non zero columns. The simplest such equations are (\ref{2sl1}) and (\ref{2sl2}).
Manipulating them, for column indices $j$ and $l$ not in $J_0$, yields:
\be
(R_{jl}-\r_{jl})\,(G_{jl}-G_{lj})=0 \hskip 5cm \nn\\
(R_{jl}-\r_{jl})\,((G_{jl}-\l_{jl})\,(G_{lj}+\l_{jl})-\g/2\,(G_{jl}+G_{lj}))=
\g\,(R_{jl}+\r_{jl})\,(G_{lj}-G_{jl}+\l_{jl})
\label{GR}
\ee
where $\ds \r_{jl}={{\l_{jl}+\g}\over{}\l_{jl}-\g}$.

At this point the set of solutions splits into two cases:

\noindent-case I 
\be
G_{jl}=G_{lj}+\l_{jl} \quad \mbox{and} \quad R_{jl}=\r_{jl}, \quad \mbox{that is} \quad K_{jj}={{f(\l_j+\g)}\over{f}} \,\prod_{l\neq j} {{\g}\over{\l_j-\l_l}} \label{solI}
\ee
-case II 
\be
G_{jl}=G_{lj}\quad \mbox{and} \quad R_{jl}=\r_{jl}\,{{(G_{jl}-\l_{jl})\,(G_{jl}+\l_{jl}-\g)}\over{(G_{jl}+\l_{jl})\,(G_{jl}-\l_{jl}-\g)}}.\label{solII}
\ee
Let us note that at this stage we cannot guarantee for case II that such an $R_{jl}$ actually correspond to the diagonal part of some $K$ matrix (the definition of $R$ in equation (\ref{XR}) is not necessarily invertible; as already mentioned this invertibility requires an extra zero-curvature condition on $R$). 

Plugging back the form (\ref{GijGj}) of $G_{ij}$ into equations (\ref{solI}) and (\ref{solII}) gives $\ds G_j=G+{{1}\over{2}}(\s+\s'+\e_{j_1}-\e_j)\,\l_j$ with $G$ an arbitrary function, $j_1$ the label of some non-zero column, $\s$ and $\s'$ two signs obeying the following constraints:
$\forall j\neq l, (\s+\s'+\e_{j_1}-\e_j-\e_l)^2=1$.

Note that setting $j=j_1$ in this constraint reads $\forall l \neq j_1, (\s+\s'-\e_l)^2=1$; the constraints can thus be factorized: 
$(\e_{j_1}-\e_j)\,(\e_{j_1}-\e_j+2\,(\s+\s'-\e_l))=0$.

Here again the set of solutions splits:

\noindent-either $\forall j, \e_j=\e_{j_1}$ (homogeneous case)  
and $\ds G_j=G+{{1}\over{2}}(\s+\s')\,\l_j$ with constraints $(\s+\s'-\e_{j_1})^2=1$, that is $(\s+\s')\,(\s+\s'-2\,\e_{j_1})=0$ or equivalently $\s+\s'=(1+\s'')\,\e_{j_1}$ with $\s''$ a sign, leading finally to:
\be
G_{ij}=G+{{1}\over{2}}(\e+\e')\,\l_j+{{1}\over{2}}(1+\e)\,\l_i,
\quad \mbox{with}\quad G(\l_j+\g)=G-{{1}\over{2}}(\e+\e')\,\g
\label{solH} \\
\e \; \mbox{and} \; \e' \; \mbox{being two arbitrary signs} \quad (\e=\e_{j_1} \; \mbox{and} \; \e'=\s''\,\e_{j_1}).\nn
\ee

-or there exists some $j_2$ such that $\e_{j_2}\neq \e_{j_1}$ (inhomogeneous case) then $\forall j \notin \{j_1,j_2\}, \e_{j_1}+(\s+\s'-\e_j)=0$.
Again we have to consider two options:

*either there exists some $j_3 \notin \{j_1,j_2\}$ such that $\e_{j_3}=\e_{j_1}$, then $\s+\s'=0$ and $\forall j\neq j_2, \e_j=\e_{j_1}$.

*or $\forall j \neq j_1, \e_j=-\e_{j_1}$ and $\s+\s'=-2\,\e_{j_1}$ .

In fact, these two cases merge into one, characterized by one special column labelled by $j_0$, such that $\forall j\neq j_0, \;\e_j=-\e_{j_0}$, then
$\ds \forall j, \;G_j=G-{{1}\over{2}}(\e_{j_0}+\e_j)\,\l_j$ and:
\be
G_{ij}=G-{{1}\over{2}}(\e_{j_0}+\e_j)\,\l_j+{{1}\over{2}}(1+\e_j)\,\l_i
=\left\{\begin{array}{l} \forall j\neq j_0,
\;G_{ij}=G+{{1}\over{2}}(1-\e_{j_0})\,\l_i \\
G_{ij_0}=G-\e_{j_0}\,\l_{j_0}+{{1}\over{2}}(1+\e_{j_0})\,\l_i
\end{array}\right.
\label{solIH}
\ee
with $ \forall j\neq j_0, \;G(\l_j+\g)=G$  and $G(\l_{j_0}+\g)=G+\e_{j_0}\,\g$.

However, reinserting the two forms (\ref{solH}) and (\ref{solIH}) in the equation with three distinct indices obtained from (\ref{ijkl}) letting $i=j=j_0$ rules out the inhomogeneous form (\ref{solIH}) and validates (\ref{solH}).

From (\ref{solH}) one has: $\ds G_{jl}-G_{lj}={{1}\over{2}}(1-\e')\,\l_{jl}$,
so case I (\ref{solI}) corresponds to $\e'=-1$ and case II (\ref{solII}) to $\e'=1$.
Now one can propose a complete expression both for $G$ and the diagonal part of K (already given in (\ref{solI}) for case I).
Given  $f$ and $g$ as any two meromorphic functions such that
$\forall j, \;g(\l_j+\g)=g$, one gets:

\noindent-case Ia ($\e=1, \;\e'=-1, \;g=G$) 
\be
G_{ij}=g+\l_i,
\quad \mbox{and}  \; K_{jj}={{f(\l_j+\g)}\over{f}}\,\prod_{l\neq j} {{\g}\over{\l_j-\l_l}} \label{solIa}
\ee
-case Ib ($\ds \e=-1, \;\e'=-1, \;g=G-\sum_{j}\l_j$)
\be
G_{ij}=g+\Sigma_j\;\mbox{with} \;\Sigma_j=\sum_{l\neq j}\l_l
\quad \mbox{and}  \;K_{jj}={{f(\l_j+\g)}\over{f}}\,\prod_{l\neq j} {{\g}\over{\l_j-\l_l}} \label{solIb}
\ee
-case IIa ($\ds \e=1, \;\e'=1, \;g=G+\sum_{j}\l_j$)
\be
G_{ij}=g-\Sigma_j+\l_i\;\mbox{with} \;\Sigma_j=\sum_{l\neq j}\l_l
\quad \mbox{and} \quad
R_{jl}={{\l_{jl}+\g}\over{}\l_{jl}-\g}\,{{(g-\Sigma_l+\l_l)\,(g-\Sigma_j+\l_j-\g)}\over{(g-\Sigma_j+\l_j)\,(g-\Sigma_l+\l_l-\g)}}, \nn \\
\mbox{that is} \quad K_{jj}={{f(\l_j+\g)}\over{f}}\,(g-\Sigma_j+\l_j)\,\prod_{l\neq j} {{\g}\over{\l_j-\l_l}}\hskip 3cm\label{solIIa}
\ee
-case IIb ($\e=-1, \;\e'=1, \;g=G$) 
\be
G_{ij}=g
\quad \mbox{and} \quad R_{jl}={{\l_{jl}+\g}\over{}\l_{jl}-\g}\,{{(g-\l_{jl})\,(g+\l_{jl}-\g)}\over{(g+\l_{jl})\,(g-\l_{jl}-\g)}}, \hskip 5cm \nn \\
\mbox{that is} \quad K_{jj}={{f(\l_j+\g)}\over{f}} \,\prod_{l\neq j} {{g+\l_j-\l_l}\over{\l_j-\l_l}}\label{solIIb}
\ee
Finally, we check that these four sets (\ref{solIa}, \ref{solIb}, \ref{solIIa}, \ref{solIIb}) actually verify all the required equations.

Let us now come back to the question of the zero columns.
We recall that starting from the introduction of $X$ and $R$ in~(\ref{XR}), we have restricted our analysis to the non-zero columns of $K$ ($j,l \notin J_0$); in particular any sum or product in (\ref{solIa}, \ref{solIb}, \ref{solIIa}, \ref{solIIb}) has to be understood within this set of non-zero columns; the same holds for the periodicity conditions on $g$.
We are actually able to reincorporate the indices of $J_0$ into the products appearing in the expressions of $K_{jj}$ using the following gauges:
\be\mbox{Cases Ia, Ib and IIa}\hspace{2.5cm}
f=\prod_{j_0 \in J_0,j\notin J_0}{{1}\over{\G({{\l_j-\l_{j_0}}\over{\g}})}}
\quad\mbox{yields}\quad
{{f(\l_j+\g)}\over{f}}=\prod_{j_0 \in J_0}{{\g}\over{\l_j-\l_{j_0}}}
\nn\\
\mbox{Case IIb}\hspace{.5cm}
f=\prod_{j_0 \in J_0,j\notin J_0}{{\G({{g+\l_j-\l_{j_0}}\over{\g}})}\over{\G({{\l_j-\l_{j_0}}\over{\g}})}}
\quad\mbox{yields}\quad
{{f(\l_j+\g)}\over{f}}=\prod_{j_0 \in J_0}{{g+\l_j-\l_{j_0}}\over{\l_j-\l_{j_0}}}.\hspace{3cm}
\nn\ee
These $J_0$-indices can then be reintroduced in $\Sigma_j$ (cases Ib and IIa), adding $\ds \sum_{l\in J_0}\l_l$ to $\Sigma_j$, up to a redefinition of $g$, which is compatible with the periodicity conditions of $g$ since they are  only required on the variables $\{\l_{i}\},\;i\notin J_0$.
From these manipulations, one can see that any solution with zero-columns indexed by $J_0$ can be obtained from a  solution with no zero-column,
first by applying a gauge transformation (multiplication on the right by the diagonal matrix with $0$ for indices in $J_0$ and 1 for other indices) and then dropping the periodicity conditions for $g$ on the variables $\{\l_{j_0},\;\forall j_0\in J_0\}$.

Another restriction has been introduced in the course of this section to get the solutions (\ref{solIa}, \ref{solIb}, \ref{solIIa}, \ref{solIIb}). The use 
of equation (\ref{ijkj}) lead us to suppose $n\ge 3$. In fact one can directly check that these solutions still are the full set of solutions for $n=2$, the only noticeable point being that for $n=2$, $\Sigma_j=\l_i$ (for $i \neq j$) and consequently cases (a) and (b) collapse both for cases I and II.

We can now fully classify the solutions of the  constant rational  semi-dynamical reflection equations.

\section{Solutions of the constant rational semi-dynamical reflection equations}

The solutions of the  semi-dynamical reflection equations (\ref{resemidyn})
with rational constant structure matrices (\ref{abcd}) belong to four sets, defined by the following formulae up to two procedures:

- a gauge transformation (\ref{gauge}) 

- or a multiplication on the right by the diagonal gauge matrix $(d_{jj}^{J_0})$ ($\forall j\in J_0, \;d_{jj}^{J_0}=0$ and $\forall j\notin J_0, \;d_{jj}^{J_0}=1$) and the elimination of the periodicity conditions on the corresponding variables $\{\l_{j_0}\}_{j_0\in J_0}$ for $g$.

In the following formulae $g$ denotes a meromorphic function $\g$-periodic on each $\l_k$.
\be
\mbox{Case Ia:} \hspace {6.25cm}
K_{ij}={{g+\l_i}\over{g+\l_j}}\,\prod_{k\neq j} {{\g}\over{\l_j-\l_k}} 
\hspace {5cm}\label{SIa}
\\
\mbox{Case Ib:} \hspace {6.25cm} K_{ij}={{g+\Sigma_j}\over{g+\Sigma_i}}\,\prod_{k\neq j} {{\g}\over{\l_j-\l_k}} 
\hspace {5cm} \label{SIb}
\\
\mbox{Case IIa:} \hspace {3.25cm}
K_{ij}=(g-\Sigma_{ij})\,\prod_{k\neq j} {{\g}\over{\l_j-\l_k}}
\quad \mbox{with} \quad \Sigma_{ij}=\sum_k\l_k-\l_i-\l_j
\hspace {2cm} \label{SIIa}
\\
\mbox{Case IIb:} \hspace {6.25cm}
K_{ij}={{g}\over{g-\l_i+\l_j}}\,\prod_{k\neq j} {{g+\l_j-\l_k}\over{\l_j-\l_k}}
\hspace {3.3cm} \label{SIIb}
\ee
The last solution (case IIb) is equivalent to the one in~\cite{ACF}, once the fully periodic $g$ function is turned into a constant (from the view point of (\ref{resemidyn}) there is no difference between a constant and a function $\g$-periodic with respect to each variable).
It thus generates, by the quantum trace construction given in~\cite{NA2}, commuting Hamiltonians of scalar $A_n$ Ruijsenaar-Schneider model.

Solutions of cases Ia and Ib are rank-one matrices, use of which to build commuting Hamiltonians is not clear (as far as the first Hamiltonian is concerned, we recover the $g \to \infty$ limit of the (case IIb) scalar $A_n$ Ruijsenaar-Schneider model). They may have to be understood within a larger structure of type "reflection-transmission"~\cite{MSR} (still to be formulated).

Finally, solution IIa yields, by the same quantum trace construction, commuting $N$-body Hamiltonians, the first one being:
\[
H=\sum_{j\in\{1\cdots N\}}\;e^{p_j}\,(g-\sum_k\l_k+2\,\l_j)\prod_{k\neq j} {{\g}\over{\l_j-\l_k}}.
\]
\section{Conclusion and prospective}  
Now that a full classification is achieved for the simplest constant
rational semi-dynamical reflection equation,
it is not to be expected that the trigonometric case yield qualitatively
different classifications of solutions. The elliptic case, if any,
is unclear: indeed dynamical constant elliptic $R$ matrices are
essentially unknown except for the algebras $A_1$ and $A_2$~\cite{BraSu}
and are known to be of a more complicated (and not explicitely
known) form for generic Lie algebras $A_n$.

The next questions to address in the
study of this algebraic structure will be the following:

\subsection{Extension to existent non-constant $ABCD$ matrices}

Known non constant rational and trigonometric $ABCD$ matrices exhibit a simple
decoupling between the constant part and the simple-pole contributions;
a fact which should be very helpful in disentangling the corresponding
reflection equations. Elliptic non-constant solutions by contrast exhibit
a coupling between spectral and dynamical parameters, but the explicit
knowledge of trigonometric solutions may help to guess the form of
some elliptic solutions. It is to be emphasized here that the Yang-Baxter
equations for $ABC$~\cite{ACF} exhibit shifts in the spectral parameters; in fact
the actual $R$ matrices constructed by Arutyunov et al.
must be defined as containing explicit differential operators, e.g
$\ds A_{12} \equiv e^{\frac{d}{dz_2}} a_{12} e^{\frac{-d}{dz_1}} $
where $a_{12}$ is the c-number matrix appearing in the reflection equation.
These shifts are related to a more general gauge covariance of the
YB and reflection equations, by l.h.s. multiplication by a 
diagonal operator involving differential operators in the spectral parameters
(used together with the gauge invariance by r.h.s. multiplication used here~(\ref{gauge})). 
Occurence of these differential operators
is not surprising here, since the derivative w.r.t. the spectral
parameter is an evaluation representation of the $d$ operator which completes the 
Cartan algebra of diagonal matrices when considering
the affine Lie algebra $\hat{A_n}$ instead of the finite Lie algebra $A_n$.
Note that the zero-weight conditions obeyed by the $ABCD$ matrices also
involve (on the explicit example of Arutyunov et al.) zero-weight conditions
for $D,B,C$ under the $d$ operator. (e.g. $D \equiv D(z_1 -z_2)$). 
Interestingly the YB equation for $D$ never acquires shifts on the
spectral parameters under these gauge transformations
due to this particular $d$-zero weight condition.

Incidentally this leads us to note that in the affine case one may add to the already
introduced dynamical variables $\lambda_i$ an extra dynamical
coordinate associated to $d$. The semi-dynamical reflection
equation may thus acquire an extra ``dynamics'' (yet to be defined)
similar somehow to the shift on the elliptic module in the affine
elliptic algebras of Jimbo et al.~\cite{JKOS}.

\subsection{Construction of new $ABCD$ sets}

The problem is to get new solutions of the YB consistency equations
with the canonical zero-weight conditions (including w.r.t. $d$ if necessary).
A possible programme for constructing (and possibly classifying) new sets
of such $ABCD$ coefficient matrices for semi-dynamical reflection equations
starts from the already known classification of 
Gervais-Neveu-Felder solutions (matrices $D$). Indeed they
are the best controlled algebraic objects, obeying the GNF equation,
with canonical zero-weight conditions and dependance on the difference
of the spectral parameters. In addition, contrary to the matrices
$A,B,C$ their YB equation is unambiguously gauge-invariant w.r.t. the left gauge multiplication, as discussed above. From a given $D$ one should then
deduce classes of $B$ and $C$ matrices, and finally $A$ matrices. It is
not clear whether the unit-solution condition $AB = CD$ must be imposed
a priori. It is not necessarily consistent with the YB equations; however
it is crucial in guaranteeing that the spin chains built from such
semi-dynamical reflection representations have a local interaction
term up to the ``dynamical'' modification, see~\cite{NA3}. Its exact
interpretation remains to be clarified, and the explicit construction
of new classes of solution may also cast some light on this problem.

\vspace {1cm}
{\bf Acknowledgements}

We wish to thank P.P. Kulish for discussion. J.A. thanks LAPTH Annecy
for their hospitality, and particularly the organizers of RAQIS 2005
for their remarkable achievement under such unhappy circumstances.
Both authors fondly remember and sorely miss our friend and collaborator 
Daniel Arnaudon.

\end{document}